\documentstyle[prl,aps,epsfig,multicol]{revtex}
\newcommand{\be}{\begin{equation}}
\newcommand{\ee}{\end{equation}}
\newcommand{\bes}{\begin{eqnarray}}
\newcommand{\ees}{\end{eqnarray}}
\newcommand{\bfig}{\begin{figure}}
\newcommand{\efig}{\end{figure}}
\newcommand{\et}{(\epsilon, \tau)}
\begin{document}      
\title{An exit-time approach to $\epsilon$-entropy} 

\author{M. Abel$^{1}$, L. Biferale$^{2}$, M. Cencini$^{1}$,
M. Falcioni$^ {1}$, D. Vergni$^{1}$ and  A. Vulpiani$^{1}$}

\address{$1$ Department of Physics and INFM, University of Rome ''La Sapienza'',\\
             piazzale Aldo Moro 2, I-00185 Roma, Italy} 
\address{$2$ Department of Physics and INFM, University of Rome ''Tor Vergata'',\\
             via della Ricerca Scientifica 1, I-00133 Roma, Italy} 
\maketitle
\vspace{0.5truecm}
\begin{abstract}
An efficient approach to the calculation of the $\epsilon$-entropy is
proposed.  The method is based on the idea of looking at the
information content of a string of data, by analyzing the signal only
at the instants when the fluctuations are larger than a certain
threshold $\epsilon$, i.e., by looking at the exit-time
statistics.  The
practical and theoretical advantages of our method with respect to the
usual one are shown by the examples of a deterministic map and a
self-affine stochastic process.
\end{abstract}


\begin{multicols}{2}
The problem of quantifying the degree of complexity of an evolving
system is ubiquitous in natural science (for a nice review see
\cite{BP97}).  The typical questions may range from the aim to
distinguish between stochastic or chaotic systems to the more
pragmatic goal of determining the degree of complexity (read
predictability) at varying the resolution in phase-space and time
\cite{GW93,Kantz}.\\ From the pioneering works of Shannon \cite{Sha48}
and Kolmogorov \cite{Kol58}, a proper mathematical tool, the
Kolmogorov-Sinai ($KS$) entropy, $h_{KS}$, has been developed to
address the above question quantitatively and unambiguously (in
principle).  The main idea is very natural: we must look at the
information contained in the time evolution as a probe of the
underlying dynamics.  This is realized by studying the symbolic dynamics
obtained by assigning different symbols to different 
cells of a finite partition of the phase space.  The
probability distribution of allowed sequences (words) is determined by
the dynamical evolution.  The average information-gain is defined by
comparing sequences of length $m$ and $m+1$, in the limit of large
$m$.  Letting the length of the words, $m$, to infinity and going to
infinitely refined partition, one obtains the $KS$-entropy, which is a
measure of the degree of complexity of the trajectory.  The $KS$-entropy
 determines also  the rate of information transmission necessary to
unambiguously reconstruct the signal.\\ Unfortunately, only in simple,
low dimensional, dynamical systems such a procedure can be properly
carried out with conventional methods~\cite{GW93,Kantz,GP83}.  The
reason is that for high dimensional systems the computational
resources are not sufficient to cope with the very high resolution and
extremely long time series required.  Moreover, in many systems, like
in turbulence, the existence of non-trivial fluctuations on different
time and spatial scales cannot be captured by the $KS$-entropy. This
calls for a more general tool to quantify the degree of predictability
which depends on the analyzed range of scales and frequencies.  This
was the aim leading Shannon \cite{shannon_eps} and Kolmogorov
\cite{Kol56} to introduce the so-called $\epsilon$-entropy, later
generalized to the $\et$-entropy \cite{GW93,GP83}.  Conceptually it
corresponds to the rate of information transmission necessary to
reconstruct a signal with a finite accuracy $\epsilon$, and with a
sampling time interval $\tau$.  The naive $(\epsilon,
\tau)$-computation is usually performed by looking at the Shannon
entropy of the coarse grained dynamics on an $(\epsilon, \tau)$-grid in
the phase-space and time.  This method  suffers of so many
computational drawbacks that it is almost useless for many realistic
time-series \cite{Kantz}. Another attempt in this direction is the
introduction of the Finite Size Lyapunov Exponent \cite{ABCPV96}.\\
The aim of this letter is to introduce an alternative approach for
the determination of the $(\epsilon, \tau)$-entropy, based on the
analysis of exit-times.  In a few words, the idea consists in looking
at the information-content of a string of data, without analyzing the
signal at any fixed time, $\tau$, but only when the fluctuations are
larger than some fixed threshold, $\epsilon$.  This simple observation
allows for a remarkable improvement of the computational possibility
to measure the $(\epsilon, \tau)$-entropy as will be discussed in
detail later.\\ We believe that the approach presented hereafter is
unavoidable in all those cases when either the high-dimensionality of
the underlying phase space, or the necessity to disentangle
non-trivial correlations at different analyzed time scales, leads to
the failure of the standard methods.\\ Let us just briefly recall the
conventional way to calculate the $(\epsilon,\tau)$-entropy for the
case of a time-continuous signal $ x(t) \in {\mathrm I\!R}$, recorded
during a (long) time interval $T$.  One defines an $\epsilon$-grid on the
phase-space and a $\tau$-grid on time.  If the motion is bounded, the
trajectory visits only a finite number of cells; therefore to each
subsequence of length $n\cdot\tau$ from $x(t)$ one can associate a
word of length $n$, out of a finite alphabet: $ W^n_t(\epsilon, \tau)
= \left( S_t,S_{t+\tau} , \dots, S_{t+(n-1)\tau} \right) $, where
$S_t$ labels the cell containing $x(t)$. From the probability
distribution of the above words one calculates the block entropies
$H_n (\epsilon,\tau)$:
\begin{equation}
\label{eq:2-4}
H_{n} (\epsilon,\tau) = - \sum _{ \lbrace W^{n}(\epsilon,\tau )\rbrace } 
P(W^{n}(\epsilon,\tau))  \ln P(W^{n}(\epsilon,\tau)).
\end{equation} 
 the $(\epsilon ,\tau)$-entropy 
per unit time, $h(\epsilon , \tau)$ is finally defined as:
\begin{eqnarray}
\label{eq:2-3a}
h_n(\epsilon , \tau)&=& {1 \over \tau} \lbrack H_{n+1} (\epsilon,\tau)
-H_n (\epsilon,\tau) \rbrack \; ,\\
\label{eq:2-3b}
h(\epsilon , \tau) &=& \lim _{n \to \infty} h_n(\epsilon , \tau) = 
{1 \over \tau} \lim _{n \to \infty} {1 \over n} H_{n} (\epsilon,\tau)\; ,
\end{eqnarray} 
where for practical reasons the dependence on the details of the partition
is ignored, while the rigorous definition is given in terms of the
infimum over all possible partitions with elements of diameter smaller
than $\epsilon$ \cite{GW93}. Notice that the above  defined
$(\epsilon,\tau)$-entropy is nothing but the Shannon-entropy 
of the sequence of
symbols $\left( S_t, S_{t+\tau}, \dots \right)$ associated with the
given signal.  The Kolmogorov-Sinai entropy, $h_{KS}$, is obtained by
taking the limit $(\epsilon,\tau) \to 0$:
\begin{equation}
\label{eq:2-5}
h_{KS} = \lim _{\tau \to 0} \lim_{\epsilon \to 0} h(\epsilon, \tau). 
\end{equation} 
In the case of discrete-time systems, one can define
$h(\epsilon)\equiv h(\epsilon,\tau=1)$, and $h_{KS} = \lim_{\epsilon
\to 0} h(\epsilon)$.  In continuous-time evolutions, whose
realizations are continuous functions of time, the $\tau$ dependence
disappears from $h(\epsilon , \tau)$, so one can still define an
$\epsilon$-entropy per unit time $h(\epsilon)$. In particular, also in
a pure deterministic flow one can put $h(\epsilon) =
h(\epsilon,\tau=1)$.\\ Let us remind that for a genuine deterministic
chaotic system one has $0< h_{KS} < \infty $ ($h_{KS}=0$ for a regular
motion), while for a continuous random process $h_{KS}=\infty$.
Therefore, in order to distinguish between a purely deterministic system 
and a  stochastic system it is necessary to perform the limit $\epsilon
\rightarrow 0$ in (\ref{eq:2-5}).  Obviously, from a physical and/or
numerical point of view this is impossible.  Nevertheless, by looking
at the behavior of the $\et$-entropy at varying $\epsilon$ one can
have some qualitative and quantitative insights on the chaotic or
stochastic nature of the process.  For most of the usual stochastic
processes one can explicitly give an estimate of the entropy scaling
behavior when $\epsilon \to 0$ \cite{GW93}.  For instance, in the case
of a stationary Gaussian process one has~\cite{GW93}
\begin{equation}
\label{eq:2-6}
h(\epsilon) \equiv \lim_{\tau\to 0} h(\epsilon,\tau) 
\sim {1 \over \epsilon^2} \; .
\label{eq:kolmo56}
\end{equation} 
Let us now introduce the main point of this letter by discussing in
detail the difficulties that may arise in measuring the $\epsilon$-entropy
for the following non-trivial example of a chaotic-diffusive  map \cite{diff}, 
\begin{equation}
x_{t+1}=x_t + p\sin 2\pi x_t\,.
\label{eq:mappa}
\end{equation}
 When $p > 0.7326\dots$, this map produces a diffusive behavior on
large scales, so one expects
\begin{equation}
\label{eq:3-2}
h(\epsilon) \simeq \lambda \,\,\,{\rm for} \,\,\, \epsilon < 1;\;\;\;
h(\epsilon) \propto {D\over \epsilon ^2} \,\,\, {\rm for} 
\,\,\, \epsilon > 1 , 
\end{equation} 
where $\lambda$ is the Lyapunov exponent and $D$ is the diffusion
coefficient.  The numerical computation of $h(\epsilon)$, using the
standard codification, is highly non-trivial already in this simple
system.  This can be seen by looking at Fig.~1 where the behavior
(\ref{eq:3-2}) in the diffusive region is roughly obtained by
considering the envelope of $h_n(\epsilon,\tau)$ evaluated for
different values of $\tau$; while looking at any single (small) value
of $\tau$ (one would like to put $\tau=1$) one obtains a rather
inconclusive result.  This is due to the fact that one has to consider
very large block lengths $n$ when computing $h\et$, in order to obtain
a good convergence for $H_{n}\et -H_{n-1}\et$ in
(\ref{eq:2-3b}). Indeed, in the diffusive regime, a simple dimensional
argument shows that the characteristic time of the system is
$T_\epsilon \approx \epsilon^2 / D$.  If we consider for example,
$\epsilon = 10$ and typical values of the diffusion coefficient $D
\simeq 10^{-1}$, the characteristic time, $T_{\epsilon}$, is much
larger than the elementary sampling time $\tau=1$.
\begin{figure}
\epsfxsize=8.5truecm
\epsfysize=6truecm
\epsfbox{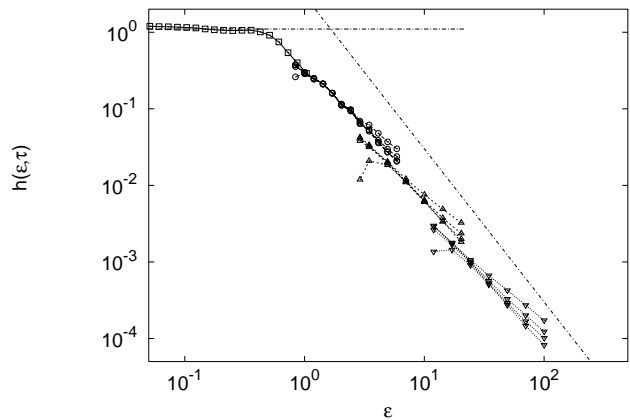}
\narrowtext
\caption{ Numerically evaluated $(\epsilon,\tau)$-entropy
for the map (\ref{eq:mappa}) with $p=0.8$ computed in the usual way
$[6]$ at $\tau=1$ ($\circ$), $\tau=10$ ($\bigtriangleup$) and $\tau=100$ (
$\bigtriangledown$) and different
block length ($n=4,8,12,20$).
The boxes ($\Box$) give the entropy computed with $\tau=1$ by
using periodic boundary condition over $40$ cells. The latter is
necessary in order to compute the Lyapunov exponent
$\lambda=h_{KS}=1.15$.  
The straight lines correspond to the two
asymptotic behaviors, $h(\epsilon)=h_{KS}$ and $h(\epsilon) \sim
\epsilon^{-2}$. 
}
\end{figure}
Our approach to calculate $h(\epsilon)$ differs from the usual one in
the procedure to construct the coding sequence of the signal at a
given level of accuracy.  Specifically, we use a different way to
sample the time, i.e., instead of using a constant time interval,
$\tau$, we sample according to the exit-time, $t(\epsilon)$, on an
alternating grid of cell size $\epsilon$.  We consider the original
continuous-time record $x(t)$ and a reference starting time $t=t_0$;
the subsequent exit-time, $t_1$, is then defined as the first time
necessary to have an absolute variation equal to $\epsilon/2$ in
$x(t)$, i.e., $|x(t_0 + t_1)-x(t_0)| \ge \epsilon/2 $. This is the
time the signal takes to exit the cell of size $\epsilon$.  Then we
restart from $t_1$ to look for the next exit-time $t_2$, i.e., the
first time such that $|x(t_0 + t_1 + t_2)-x(t_0 + t_1)| \ge \epsilon/2
$ and so on.  Let us notice that, with this definition, the
coarse-graining grid is not fixed, but it is always centered in the
last exit position.  In this way we obtain a sequence of exit-times,
$\{t_i(\epsilon)\}$.  To distinguish the direction of the exit (up or
down out of a cell), we introduce the label $k_i= \pm 1$, depending
whether the signal is exiting above or below. Doing so, the trajectory
is univocally coded with the required accuracy, by the sequence
$((t_1,k_1), (t_{2},k_{2}), \dots , (t_{M},k_{M}))$, where $M$ is the
total number of exit-time events observed during the total time $T$.
Correspondingly, an ``exit-time word'' of length $n$ is a sequence of
couples of symbols $\Omega^n_i(\epsilon)=((t_i,k_i),
(t_{i+1},k_{i+1}), \dots , (t_{i+n-1},k_{i+n-1}))$.  From these words
one firstly calculates the block entropies, $H^\Omega_n(\epsilon)$,
and then the exit-time $\epsilon$-entropies: $h^\Omega(\epsilon)
\equiv \lim_{n \to \infty} H^\Omega_{n+1}(\epsilon) -
H^\Omega_n(\epsilon)$.  Let us notice that $h^\Omega(\epsilon)$ is an
$\epsilon$-entropy {\it per} exit and that $M = T/{\langle t(\epsilon)
\rangle}$. The exit-time coding is a faithful reconstruction with
accuracy $\epsilon$ of the original signal. Therefore, the total
entropy, $M h^\Omega(\epsilon)$, of the exit-time sequence,
$\Omega^M(\epsilon)$, is equal to the total entropy, $T h(\epsilon) =N
\tau h(\epsilon)$, of the standard codification sequence,
$W^N(\epsilon)$.  Namely, for the $\epsilon$-entropy per unit time, we
obtain:
\begin{equation}
h(\epsilon) = M  h^\Omega(\epsilon) / T =
\frac{h^\Omega(\epsilon)}{\langle t(\epsilon) \rangle}\,\,.
\label{epsent}
\end{equation}
Now we are left with the determination of $h^\Omega(\epsilon)$.  This
implies a discretization, $\tau_e$, of the exit-times.  The
exit-time entropy $h^\Omega(\epsilon)$ becomes an
exit-time $(\epsilon,\tau_e)$-entropy, 
$h^\Omega(\epsilon,\tau_e)$, obtained
from the sequence $\{ \eta_i,k_i \}$, where $\eta_i$ identifies the
exit-time cell containing $t_i$.  Equation (\ref{epsent}) becomes now
\begin{equation}
h(\epsilon) =\lim_{\tau_e \to 0}
h^\Omega(\epsilon,\tau_e)/\langle t(\epsilon)\rangle \simeq
h^\Omega(\epsilon,\tau_e) / \langle t(\epsilon) \rangle \,\,,
\label{ep_master}
\end{equation}
the latter relation being valid for small enough $\tau_e$.  However,
in all practical situations there exist a minimum $\tau_e$ given by
the highest acquisition frequency, i.e., the limit $\tau_e \to 0$
cannot be reached. The discretization interval $\tau_e$ can be thought
as the equivalent to the $\tau$ entering in the usual $\et$-entropy,
so that
\begin{equation}
h(\epsilon,\tau_e) = h^\Omega(\epsilon,\tau_e) / \langle t(\epsilon)
\rangle \,\,.
\label{bho}
\end{equation}
At this point it is important to stress that in most of the cases  the leading
$\epsilon$ contribution to $h(\epsilon)$ in (\ref{ep_master}) is given
by the mean exit-time $\langle t(\epsilon) \rangle$ and not by
$h^\Omega(\epsilon,\tau_e)$. Anyhow,  the computation
of $h^\Omega(\epsilon,\tau_e)$ is compulsory in order  to recover
a zero entropy for regular (e.g. periodic) signals.
It is easy to obtain the following bounds for 
$h^\Omega(\epsilon,\tau_e) = h^\Omega(\{ \eta_i,k_i \})$~:
$$h^\Omega(\{ k_i \}) \leq h^\Omega(\{ \eta_i,k_i \})$$
$$ h^\Omega(\{ \eta_i,k_i \})\leq h^\Omega(\{ \eta_i\}) +
h^\Omega(\{k_i\}) \leq h^\Omega(\{ k_i \}) + H^\Omega_1(\{\eta_i\})$$
where $h^\Omega(\{k_i\})$ is the Shannon entropy of the sequence
$\{k_i\}$ and $H^\Omega_1(\{\eta_i\})$ is the one-symbol entropy of
the $\{ \eta_i\}$.  Therefore we have
\begin{equation}
\label{bound-entro}   
{h^\Omega(\{ k_i \}) \over \langle t(\epsilon) \rangle} \leq 
h(\epsilon) \leq 
{h^\Omega(\{ k_i \}) + c(\epsilon) + \ln (\langle t(\epsilon)\rangle / \tau_e)
\over \langle t(\epsilon) \rangle}
\end{equation} 
where $c(\epsilon) = -\int p(z)\ln p(z) {\mathrm d}z$, and $p(z)$ 
is the probability distribution function of the rescaled exit-time 
$z(\epsilon) = t(\epsilon)/\langle t(\epsilon)\rangle$. \\
We shall see in the following that the above bounds are rather good, and
typically $\langle t(\epsilon) \rangle$ shows the same scaling
behavior as $h(\epsilon)$.  One could wonder why the exit-time
approach is better than the usual one.  The reason is simple (and
somehow deep): in the exit-time approach it is not necessary to use a
very large block size since, at fixed $\epsilon$, the typical time at
that scale is automatically given by $ \langle t(\epsilon) \rangle $.
This fact is particularly clear in the case of Brownian motion.  In
such a case $ \langle t(\epsilon) \rangle \propto \epsilon ^2 / D$,
where $D$ is the diffusion coefficient.  As previously discussed, the
computation of the $h(\epsilon)$ with the standard methods implies the
use of very large block sizes, of order $\epsilon ^2 / D$.
\begin{figure}
\epsfxsize=8.5truecm
\epsfysize=6truecm
\epsfbox{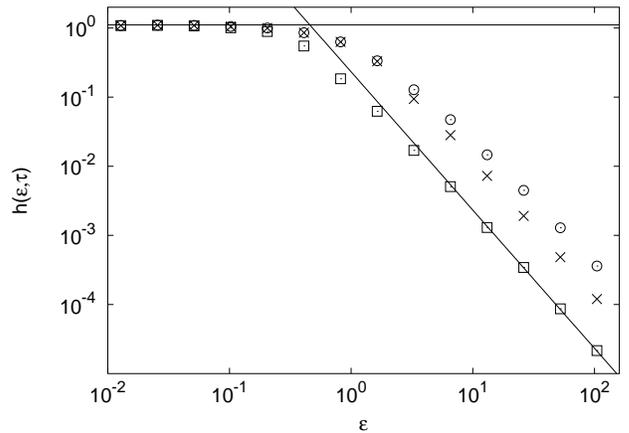}
\narrowtext
\caption{Numerically computed lower bound ($\Box$) and 
upper bound (with $\tau_e=1$)
 ($\circ$)
of $h(\epsilon)$ (\ref{bound-entro}) for the same map of Fig.~1.
The two straight (solid) lines correspond
 to the asymptotic behaviors as in Fig.~1. 
We also present the $(\epsilon,\tau_e)$-entropy 
$h^\Omega(\epsilon,\tau_e)/\langle t(\epsilon)\rangle$
with $\tau_e=0.1 \langle t(\epsilon)\rangle$ ($\times$).}
\end{figure}
With our method the $\langle t(\epsilon) \rangle$ captures the correct
scaling behavior and the exit-time entropy introduces, at worst, a
sub-leading logarithmic contribution $h^\Omega(\epsilon,\tau_e) \sim
\ln (\langle t(\epsilon) \rangle /\tau_e)$. This is because
$c(\epsilon)$ is $O(1)$ and independent of $\epsilon$ for a
self-affine signal and the $h^\Omega(\{ k_i \}) \leq \ln 2$ term is
small compared with $\ln (\langle t(\epsilon) \rangle /\tau_e)$ (for
not too small $\epsilon$), so that, neglecting the logarithmic
corrections, $h(\epsilon) \simeq 1/\langle t(\epsilon) \rangle \propto
D \epsilon^{-2}$. In order to clarify this point, we plot in Fig.~2
the calculation of the $\et$-entropy {\it via} the exit-time approach
for the previously discussed diffusive map.  Fig.~2 has to be compared
with Fig.~1 where the usual approach has been used.  While in Fig.~1
the expected $\epsilon$-entropy scaling is roughly recovered as an
envelope over many different $\tau$, in our case the predicted
behavior is easily recovered for any $\epsilon$ with a remarkable
improvement in the quality of the result.\\ Let us now briefly comment
the limit $\epsilon \to 0$ for a discrete-time system (e.g. maps). In
this limit the exit-time approach coincides with the usual one: we
have just to observe that practically the exit-times always coincide
with the minimum sampling time and to consider the possibility to have
jumps over more than one cell, i.e., the $k_i$ symbols may take values
$\pm 1,\pm 2, \dots$.
\begin{figure}
\epsfxsize=8.5truecm
\epsfysize=6truecm
\epsfbox{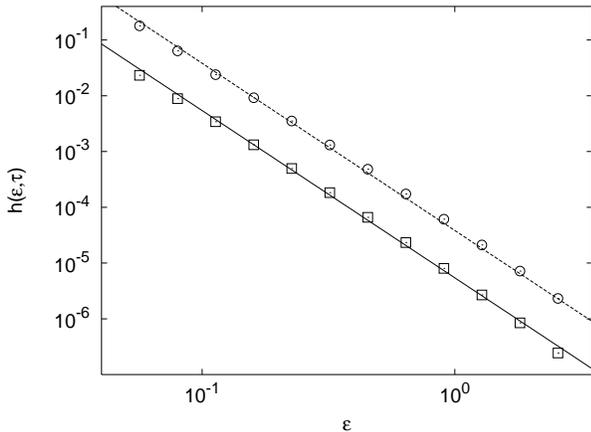}
\narrowtext
\caption{Numerically computed lower bound ($\Box$) 
and upper bound (with $\tau_e=0.1\langle t(\epsilon)\rangle$) ($\circ$)
for the $(\epsilon,\tau_e)$-entropy in the case
of a self-affine signal with $\xi=1/3$ evaluated by using the exit-time approach.
The two straight lines show the scaling $\epsilon^{-3}$.}
\end{figure}
As another example, we present the calculation of $\epsilon$-entropy
for a self-affine stochastic signal with H\"older-exponent $\xi= 1/3$,
i.e., $|x(t)-x(t+\Delta t)| \sim (\Delta t)^{1/3}$. Such a signal can
be seen as a stochastic surrogate of a turbulent signal (ignoring
intermittency) and can be constructed in different ways (see
Ref.~\cite{multiaffine} and references therein). A simple dimensional
estimate, which is rigorous for Gaussian processes~\cite{Kol56}, tells
us that the leading contribution to the $\epsilon$-entropy scaling is
given by $h(\epsilon) \sim \epsilon^{-3}$.  To generate the
self-affine signal we use a recently proposed algorithm
\cite{multiaffine}, where $x(t)$ is obtained by using many Langevin
processes.  In Fig.~3 we show the bounds (\ref{bound-entro}) for
$(\epsilon,\tau_e)$-entropy calculated {\it via} the exit-time
approach.  We observe an extended region of well-defined scaling,
which is the same for $1/\langle t(\epsilon)\rangle \sim
\epsilon^{-3}$.  The usual approach (not shown) gives a poor estimate
for the scaling as the envelope of $h\et$ computed for various $\tau$
(see for example Figs. 15-18 in \cite{GW93}), as in the case of
Fig.~1.\\ In conclusion, we have introduced an efficient method to
calculate the $\epsilon$-entropy based on the analysis of the
exit-time statistics. It is able to disentangle in a more proper way
the leading contributions to $h(\epsilon)$ at the scale $\epsilon$,
compared with the standard way based on a coarse-grained dynamics on a
fixed $\et$ grid.  We have presented the application to two examples,
a chaotic diffusive map and a stochastic self-affine signal.  More
applications to chaotic systems, stochastic multi-affine processes,
and experimental turbulent signals will be reported elsewhere.\\ The
entropy-approach to evaluate the degree of complexity of a time series
allows also to attack much more sophisticated problems often
encountered in dynamical system theory.  We just mention, e.g., the
problem to disentangle correlations between time series obtained by
measuring different observable of the same system.  This might be
addressed by using the conditional-entropy. \\ We acknowledge useful
discussions with G.~Boffetta and A.~Celani.  This work has been
partially supported by INFM (PRA-TURBO) and by the European Network
{\it Intermittency in Turbulent Systems} (contract number
FMRX-CT98-0175). M.A. is supported by the European Network {\it
Intermittency in Turbulent Systems}.

\end{multicols}
\end{document}